\documentclass{nature}

\usepackage{amsmath}                  
\usepackage{physics}
\usepackage{caption}
\usepackage{graphicx}
\usepackage{bbm}         
\usepackage{amsfonts}          
\usepackage{amssymb}            
\usepackage{mathrsfs}
\usepackage{mathtools}

\usepackage{color}

\newcommand{\ie}{\textit{i.e.\ }\/}
\newcommand{\vect}[1]{\mathbf{#1}}
\newcommand{\e}{e}
\newcommand{\im}{i}
 
\newcommand{\nvec}[1]{\vect{#1}}
\newcommand{\ee}{\e} 
\newcommand{\ii}{\im} 
\newcommand{\trans}{\top}
\newcommand{\reT}{\mathfrak{Re}\,}

\makeatletter
\let\saved@includegraphics\includegraphics
\AtBeginDocument{\let\includegraphics\saved@includegraphics}
\renewenvironment*{figure}{\@float{figure}}{\end@float}
\makeatother  

\newcommand{\ltspice}{\mbox{\textit{LTspice }}}

\title{Observation of bulk boundary correspondence breakdown in topolectrical circuits}

\author{T. Helbig$^{1}$, T.~Hofmann$^{1}$, S.~Imhof$^{2}$,
  M.~Abdelghany$^{2}$, T. Kiessling$^{2}$, L.~W.~Molenkamp$^{2}$,
  C.~H.~Lee$^{3}$, A.~Szameit$^{4}$, M.~Greiter$^{1}$ \& R.~Thomale$^{1}$}

\begin{document}

\maketitle

\begin{affiliations}
 \item Institut f\"ur Theoretische Physik und Astrophysik, Universit\"at W\"urzburg, D-97074 W\"urzburg, Germany
 \item Physikalisches Institut and R\"ontgen Research Center for
   Complex Material Systems, Universit\"at W\"urzburg, D-97074
   W\"urzburg, Germany
\item Department of Physics, National University of Singapore, Singapore, 117542
\item Institut f\"ur Physik, Universit\"at Rostock, Albert-Einstein-Str. 23, 18059 Rostock, Germany
\end{affiliations}

\begin{abstract}

The study of the laws of nature has traditionally been pursued in the limit of isolated systems, where energy is conserved.  This is not always a valid approximation, however, as the inclusion of features like gain and loss, or periodic driving, qualitatively amends these laws.  A contemporary frontier of meta-material research is the challenge open systems pose to the established characterization of topological matter\cite{brad,PhysRevX.8.031079}.  There, one of the most relied upon principles is the bulk-boundary correspondence (BBC), which intimately relates the properties of the surface states to the topological classification of the bulk\cite{PhysRevLett.71.3697,prodan2016bulk}.  The presence of gain and loss, in combination with the violation of reciprocity, has recently been predicted to affect this principle dramatically\cite{yao2018edge,Xiong_2018}.  Here, we report the  experimental observation of BBC violation in a non-reciprocal topolectric circuit\cite{leecommphys}. The circuit admittance spectrum exhibits an unprecedented sensitivity to the presence of a boundary, displaying an extensive admittance mode localization despite a translationally invariant bulk. Intriguingly, we measure a non-local voltage response due to broken BBC. Depending on the AC current feed frequency, the voltage signal accumulates at the left or right boundary, and increases as a function of nodal distance to the current feed.
\end{abstract}

The difference between a translationally invariant isolated ring of $N$ sites with periodic and open boundary conditions (PBC and OBC) is just one single bond. A generic physical system of such kind would hence be expected to exhibit an overall change of energy eigenvalues from PBC to OBC that is perturbatively small. Exceptions are given by topological edge modes that may appear at the open ends: even though the overall energy spectral flow is of the order of $1/N$ per state, it can accumulate for a few states located at the boundary. This is the principal motif of BBC in a system where the topological invariant in the PBC case directly connects to the corresponding boundary modes in the OBC case. The BBC can be violated as soon as we consider open systems in which gain and loss conspire with non-reciprocity. First hints along this line were noted by anomalous localization found in a disordered model for diffusive and convective biological processes, where growth and death terms represent gain and loss and the convective drift implies non-reciprocity\cite{hatano1996localization,nelson1998non}. Recently, similar anomalous spectral flow was theoretically noted in translationally invariant systems: The transition from PBC to OBC becomes dramatically non-perturbative, as all states  can exhibit localization at one boundary\cite{yao2018edge}. 

In order to realize this phenomenon in experiment, we design an electric circuit that represents a one-dimensional non-Hermitian non-reciprocal two-band model, as depicted in Fig. 1. Each circuit unit cell consists of two nodes $A$ and $B$ with intercell coupling $r$ and intracell coupling $v\pm\gamma$, where $\gamma$ thus represents the non-reciprocal conductance contribution (Fig. 1a). We have built a circuit chain of $N=10$ unit cells (Fig. 1b). The circuit elements of the unit cell are specified in Fig. 1c, while a physical unit cell board cutout is presented in Fig. 1d. Via the Kirchhoff rule $I=JV$, where $I$ denotes the current input and $V$ the voltage measured against ground at each node, the admittance matrix $J$ takes a block diagonal form $J(k)$ in a translationally invariant circuit network and allows to define momentum-resolved admittance bands $j(k)$ accessible through elementary impedance measurements\cite{leecommphys,quadrupolar,helbigprb}. Up to prefactors, the admittance matrix $J$ and its eigenvalues $j(k)$ take a role similar to a Hamiltonian and its band structure of energy eigenvalues.
While capacitive and inductive elements represent Hermitian couplings, gain is realized by operational amplifiers and loss is given by serial resistances. Furthermore, operational amplifiers arranged as impedance converters through current inversion\cite{PhysRevLett.122.247702} allow to precisely tune the type of non-reciprocity in the circuit (see Methods section). For a given AC input current of frequency $\omega=2\pi f$, we eventually arrive at the non-reciprocal two-band admittance model
\begin{align}\label{eq:Laplacian}
(i \omega)^{-1} \, J (k) = \epsilon_0 (\omega)\, \mathbbm{1}+
\Big[v(\omega) + r \cos(k)\Big] \, \sigma_x
+\Big[ r  \sin(k) - i  \gamma \Big] \, \sigma_y,
\end{align}
where $\sigma_{x,y}$ denote Pauli matrices and the parametric functions are given by $\epsilon_0(\omega) = C_1+C_2+C_3-\frac{1}{\omega^2  L_0 }-\frac{1}{\omega^2 L_1 } -\frac{i}{\omega R_0 }$, $v(\omega) = C_1 -\frac{1}{\omega^2 L_1}$, $r=C_2$, and $\gamma=C_3$. Non-Hermiticity in (1) is hence accomplished by asymmetric off-diagonal couplings. The real and imaginary part of the admittance spectrum (1) is plotted in Fig. 2a for OBC and PBC, respectively. Note that $\omega$ (and accordingly $f$) is an external parameter of the circuit model which can be tuned at will. The crucial challenge for the experimental design of (1) is to accomplish circuit stability in the combined presence of parasitic effects and gain elements. This was reached by a concerted effort of circuit element disorder analysis and refined circuit board architecture (see Methods section).
$v(\omega)$ can change its sign as a function of $\omega$, and fundamentally modifies the admittance spectrum (Fig. 2a). At $\abs{v(\omega)}=\abs{\gamma}$, which occurs at the frequencies $f_\text{EP,1}= \left(2\pi \sqrt{ (C_1+C_3) \, L_1}\right)^{-1}\approx$ 85.0\,kHz and $f_\text{EP,2}= \left(2 \pi \sqrt{(C_1-C_3) \, L_1}\right)^{-1}\approx$ 99.6\,kHz, the circuit features a spectral singularity commonly referred to as exceptional point. There, multiple eigenvalues and their corresponding eigenvectors coalesce, a defective spectral feature which cannot appear in a Hermitian system. Exceptional points have been previously observed and addressed for non-Hermitian models realized in photonic crystals\cite{zhen2015spawning, chen2017exceptional,aash,Mirieaar7709}. The exceptional points are visible as admittance bifurcation points for OBC and PBC in Fig. 2a.

In Fig. 2b we plot the measured admittance spectrum at five representative values (i)-(v) of $f$, as highlighted in Fig. 2a, for PBC and OBC along with the voltage eigenstates for OBC. PBC (OBC)  is accomplished by (not) connecting node 1 with 20 (Fig. 1b). The measured data of the $N=10$ chain is denoted by points, in excellent agreement with the aberration-corrected theoretical data highlighted by lines (see Methods section). The OBC spectrum looks drastically different from the PBC spectrum, stressing the non-perturbative spectral flow. While the PBC eigenspectrum traces out closed loops in the complex admittance plane, these loops are deformed to open arcs or points under the spectral flow evolution to the OBC spectrum. For the OBC case, all voltage eigenmodes localize at one boundary. Within our frequency sweep, we pass from localization at the left boundary for $v(\omega)<0$ over a transitional point with delocalized modes in Fig. 2b(iii) at $v(\omega)=0$ corresponding to $f=\left(2 \pi \sqrt{L_1\,C_1} \right)^{-1}\approx$ 91.4\,kHz to localization at the right boundary for $v(\omega)>0$. The strongest localization is found around the two exceptional points for $v(\omega)=- \gamma$ and $v(\omega)= +\gamma$ depicted in Fig. 2b(ii) and Fig. 2b(v), respectively. For OBC, we further find topological Su-Schrieffer Heeger midgap states\cite{leecommphys} for $\vert v \vert < \sqrt{r^2+\gamma^2}$, which does not match the band closing points $\vert v \vert=\vert r\pm \gamma \vert$ inferred from the PBC spectrum. This shows that firstly, topological modes can coexist with the localization of all bulk modes, and secondly, topological phase transitions expected from a bulk analysis are parametrically distorted due to non-Hermiticity\cite{lee2016anomalous,kunst2018biorthogonal,PhysRevB.99.201103}.
In Fig. 2c, we analyze the admittance spectral flow as we interpolate from PBC to OBC. Starting from our PBC circuit configuration, this is done by continously attenuating the 1-20 bond to zero conductance. The interpolation is quantified by $\eta$, which is the 1-20 bond capacitance normalized to its PBC value. This observation explicates the remarkable sensitivity of the admittance spectrum with respect to a single bond attenuation.

BBC breakdown manifests itself as a non-local voltage response in our circuit. As displayed in Fig. 3, we inject the AC current feed at different nodes on the left side of our circuit chain. In a regular passive circuit array, the principal notion of locality would suggest a voltage profile with predominant weight around the location of the current feed.
Instead, for an AC frequency $f$ such that $v>0$, our circuit produces a dominant voltage signal at the right edge, which is even enhanced the further the current feed is away from the right end (inset Fig. 3). This is a consequence of extensive bulk mode localization at the right edge. Most importantly, this accumulation at the right edge can be tuned into an accumulation at the left edge just by changing $f$, and as such the sign of $v$ in (1).

The breakdown of bulk boundary correspondence, as observed in our electric circuit array with a non-reciprocal non-Hermitian admittance profile, relates to a drastic admittance spectral flow from periodic to open boundary conditions. It implies the localization of all bulk modes, and furthermore displays a tunable non-local voltage response to a local current feed. In perspective of contemporary technological directions of electrical engineering, future studies could advance the intertwining of non-linear circuitry and the peculiar open system effects studied here\cite{dunkel,hadad}.  Furthermore, the insights gained from our topolectrical circuit readily promise a transfer to other platforms such as photonic, mechanical, acoustic, or other metamaterial settings~\cite{coulais,wang}.

\bibliography{paper-skin}

\begin{addendum}
 \item[Acknowledgments] The circuit simulations have been performed by the use of
   \ltspice. The work in W\"urzburg is
   funded by the Deutsche Forschungsgemeinschaft (DFG, German Research
   Foundation) through Project-ID 258499086 - SFB 1170 and through the
   W\"urzburg-Dresden Cluster of Excellence on Complexity and Topology
   in Quantum Matter -- \textit{ct.qmat} Project-ID 39085490 - EXC
   2147.
\item[Author contributions] T.He., T.Ho., C.H.L., and R.T. conceptualized the
   project, all authors designed the experimental setup and
   theoretical modelling. S.I., M.A., and T.K. carried
   out the circuit experiments and data analysis, T.He. and T.Ho. performed
   the \ltspice simulations. R.T. supervised the investigation and
   wrote the paper with key contributions from T.He., T.Ho., and
   C.H.L. The manuscript reflects the contributions of all authors.
 \item[Competing Interests] The authors declare no competing financial interests. 
 \item[Correspondence] Correspondence and requests for materials should be addressed to R.T. (rthomale@physik.uni-wuerzburg.de).
\end{addendum}

{\bf Caption Figure 1:} {\bf Reciprocity breaking in a non-Hermitian linear circuit. (a)} Asymmetric intracell couplings $v \pm \gamma$ imply a non-Hermitian non-reciprocal tight-binding model. {\bf (b)} Sketch of the total circuit configuration consisting of $N=10$ unit cells for periodic (red) and open (green) boundary conditions. {\bf (c)} Experimental implementation of a non-reciprocal, non-Hermitian circuit model based on {\bf (a)}. The $\gamma$ term is achieved via negative impedance converters with current inversion (INICs). A resonating $LC$ circuit in the intracell coupling tunes $v(\omega)$ in (1). {\bf (d)} Circuit board cutout of one unit cell.

{\bf Caption Figure 2:} {\bf Bulk-boundary correspondence breakdown in the admittance spectrum.  (a)} Admittance spectrum of (1) as a function of external AC frequency $f$ split into imaginary and real part for PBC and OBC. Line cuts (i) to (v) highlight representative values of $f$ which are further discussed in {\bf b}.  {\bf (b)} Complex admittance for AC frequencies $f_{\text{(i)}}=70$\,kHz, $f_{\text{(ii)}}=84.2$\,kHz, $f_{\text{(iii)}}=91.5$\,kHz, $f_{\text{(iv)}}=95$\,kHz, and $f_{\text{(v)}}=98.5$\,kHz of the circuit model with PBC in red and OBC in green. Discrete dots denote experimental data, dashed and joint lines show the parasitics-corrected theoretical band predictions. The Su-Schrieffer-Heeger edge modes (blue crosses) appear for (ii)-(v). 
For $v\neq 0$, the OBC spectrum differs drastically from the PBC spectrum. For each frequency $f$, voltage amplitudes $\abs{\psi_n}$ are shown for the $n$th eigenvector with OBCs as a function of the node index $x$. As seen in (i), (ii)  ((iv), (v)), for $v<0$ ($v>0$), all voltage modes localize at the left (right) circuit boundary. The itinerant transition point is reached for $v=0$ in (iii), where no bulk mode localization is observed.
{\bf (c)}  Spectral flow of the admittance spectrum from PBC to OBC for $f=80$ kHz, as the 1-20 bond is continously modified from PBC ($\eta=1$) to OBC ($\eta=0$). Experimental data is shown as black dots combined with the predicted PBC and OBC spectrum in red and green, respectively.

{\bf Caption Figure 3:} {\bf Non-local voltage response.} A current feed is imposed on the left side of the circuit realized through a voltage source of AC driving $f=98.5$ kHz connected to the circuit through a shunt resistance $R_\text{S} = 12.0\; \Omega$. The main voltage response to the left side current feed is centered on the right edge of the system. The inset shows the maximum amplitude $V_{\text{max}}$ with respect to the voltage level at the feed node against the nodal position of the current feed. $V_{\text{max}}$ increases as a function of the nodal distance of the current feed to the right edge. The side at which the voltage pulse localizes can be switched by $f$.

\clearpage
\begin{figure}
\centering
\includegraphics[width=0.9\linewidth]{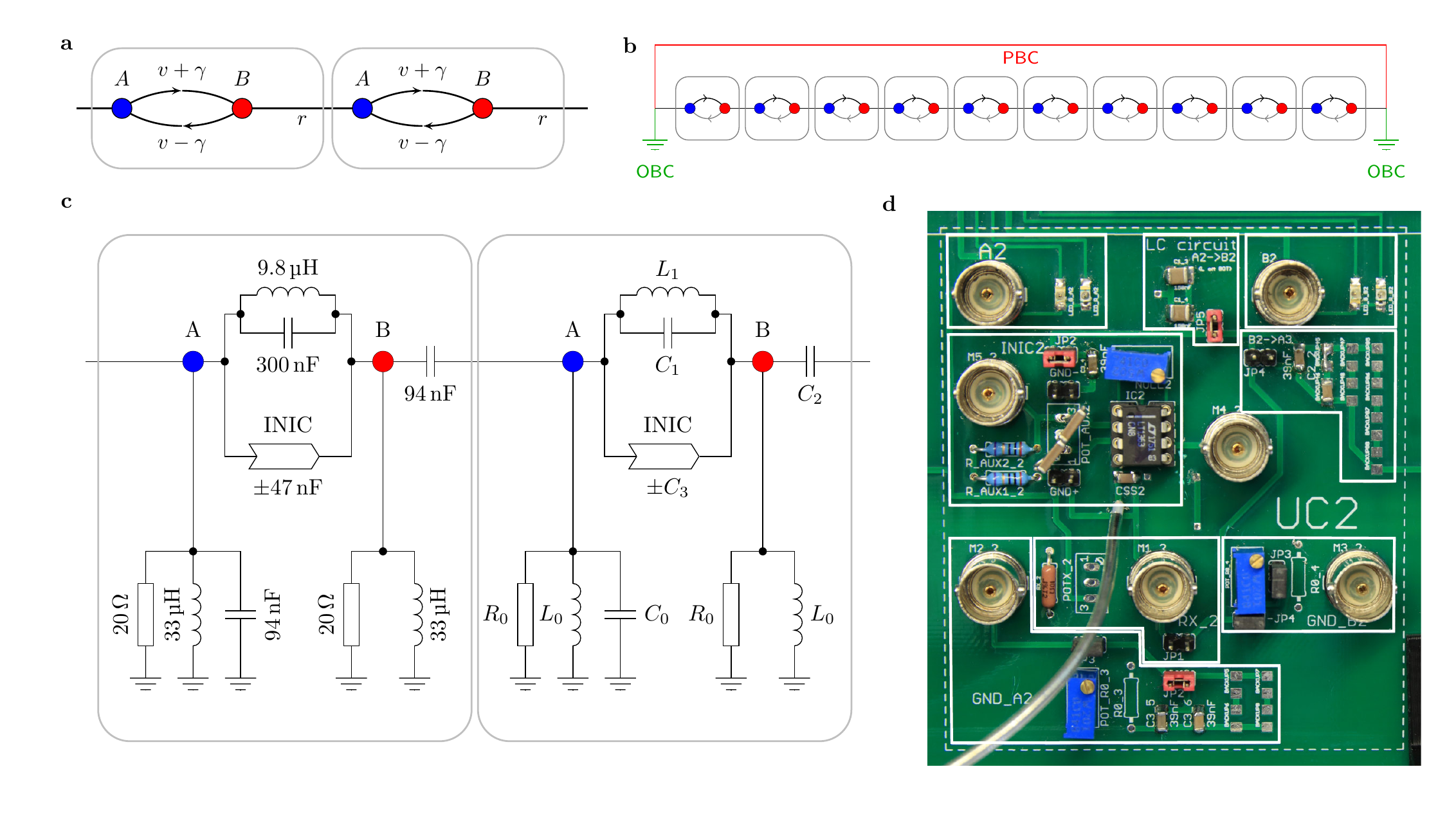}
\caption{}
\end{figure}

\begin{figure}
\centering
\includegraphics[width=\linewidth]{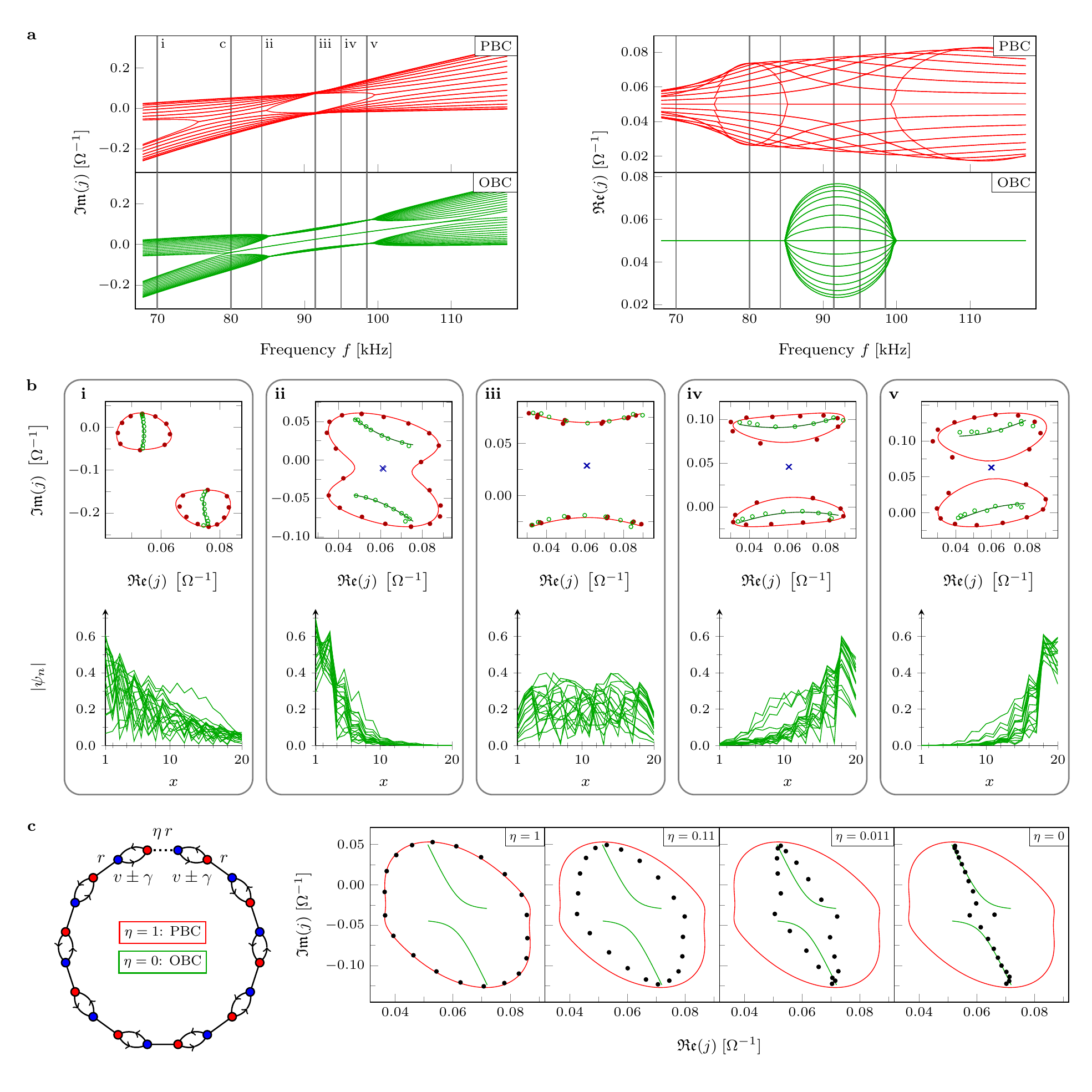}
\caption{}
\label{fig:eigenvalues}
\end{figure}

\begin{figure}
\centering
\includegraphics[width=\linewidth]{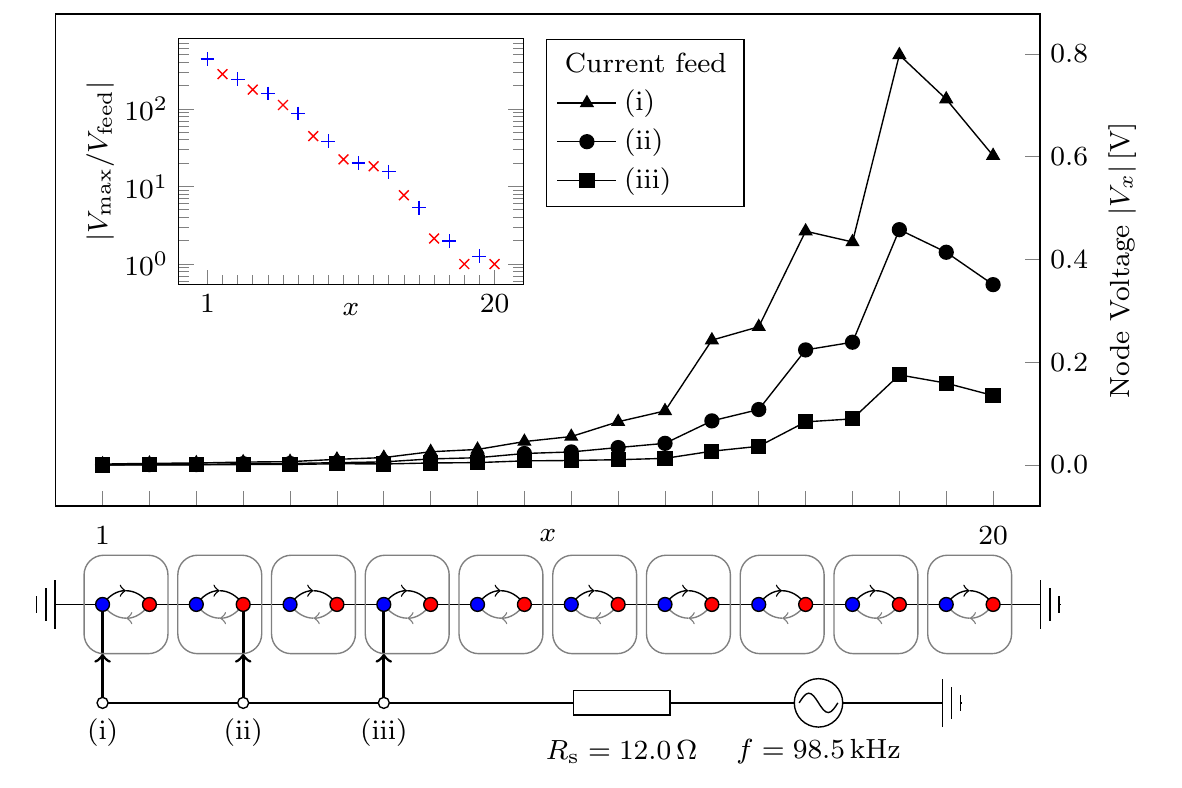}
\caption{}
\end{figure}

\clearpage
\begin{center}
	\LARGE{METHODS}
\end{center}

\section{Topolectrical circuits design}

\begin{figure}
	\centering
	\includegraphics[width=\linewidth]{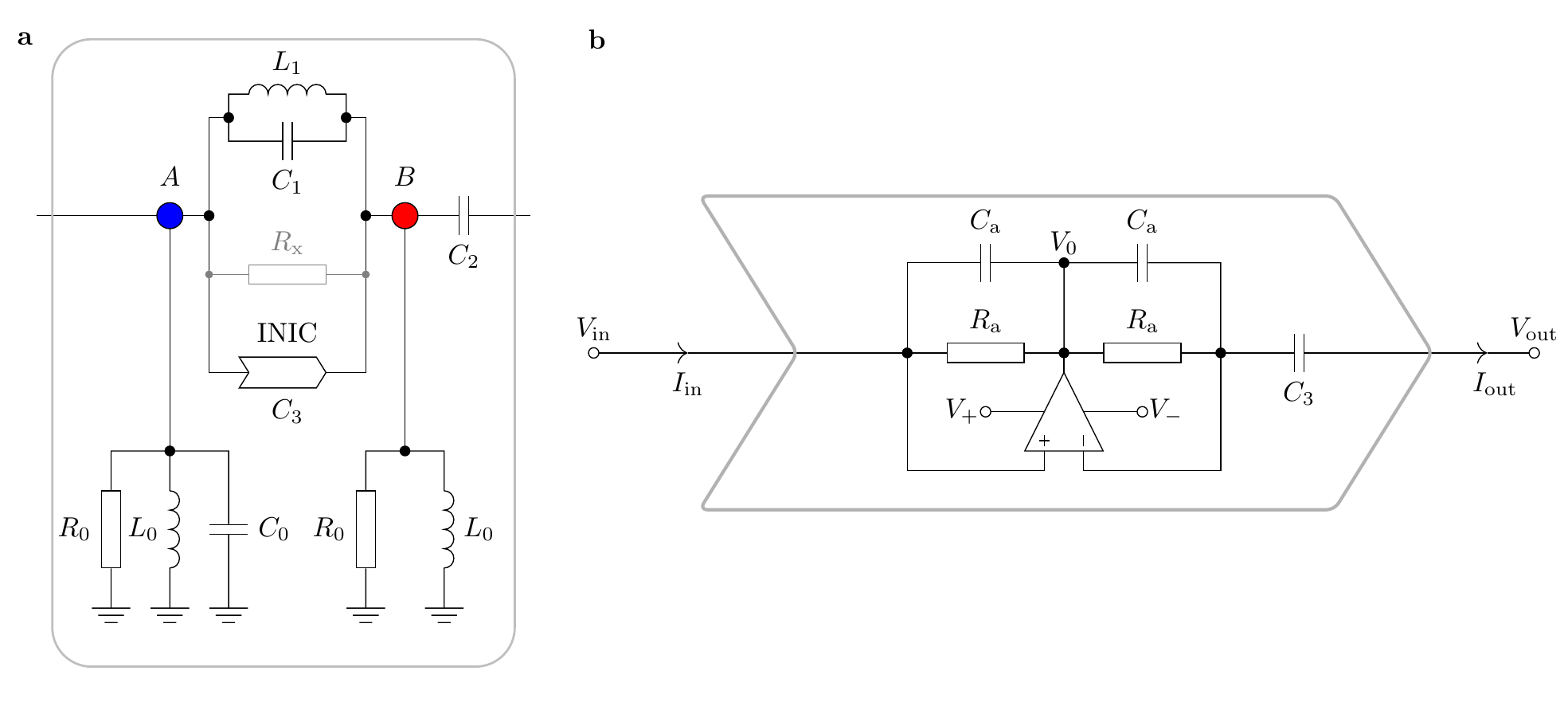}
	\caption{}
	\label{fig:circuit_schematic_simplified}
\end{figure}

We design a circuit that implements both non-reciprocal and non-Hermitian couplings in a periodical network. Topological effects in electrical circuit networks have initially been considered for the Hofstadter model~\cite{PhysRevX.5.021031,PhysRevLett.114.173902}. The Laplacian formalism, with the complex admittance as its observable, has been introduced in~\cite{leecommphys}. It has established the notion of admittance band structures, which are fully accessible experimentally~\cite{helbigprb}. Topolectrical circuits are a newly emerging platform for synthetic topological matter which allows to implement any tight-binding model of any coupling range or dimensionality. As such, topolectric circuits manage to host nodal knots or links as zero admittance eigenvalues in their three-dimensional variant~\cite{Research2018Yu,lee2019topological}. Topological corner states as implementations of higher order topology can likewise be realized~\cite{quadrupolar,PhysRevB.100.045407}. 
Our non-Hermitian non-reciprocal circuit network is described by the model Laplacian introduced in Eq.~(1). An idealized conceptual unit cell of the circuit is shown in Fig.~\ref{fig:circuit_schematic_simplified}a. The unit cell consists of two sublattice nodes, $A$ and $B$. The sublattices are connected in the unit cell by an $LC$-resonating circuit, which represents the Hermitian and reciprocal coupling $\ii \omega\,v(\omega) = \ii \omega C_1 + 1/(\ii \omega L_1) $. The frequency-dependence of its impedance allows us to modulate the intracell coupling and to resolve the fundamentally different parameter regimes of the model on the same circuit board. The $LC$ oscillator is characterized by its resonance frequency $\omega_0^{-1} = \sqrt{L_1 C_1}$, at which any current is blocked from flowing	through the component by an impedance divergence, such that the connectivity $v(\omega_0) = 0$. The coupling $C_2$ in between unit cells acts as the reciprocal intercell connection $\ii\omega \, r = \ii \omega C_2$.
Inevitable resistive losses in the circuit can, to first approximation, be modeled by an effective resistance $R_\text{x}$ connecting the nodes $A$ and $B$ within an unit cell. %This contribution, though time-reversal symmetry breaking and non-Hermitian, cannot induce a non-Hermitian skin effect since it is not direction-modulated and therefore non-reciprocal.~\cite{SuppMat}

The key non-Hermitian component in our circuit is the negative impedance converter with current inversion (INIC), which implements non-reciprocity~\cite{PhysRevLett.122.247702}. Its setup is shown in Fig.~\ref{fig:circuit_schematic_simplified}b. Through an operational amplifier (OpAmp) in a negative feedback configuration, we realize an antisymmetric connection, that implements a positive capacitance $C_3$ in one direction and a negative capacitance with equal magnitude in the opposing direction, leading to the breaking of reciprocity. Solving Kirchhoff's laws for the INIC leads to the reduced two-node Laplacian
\begin{align}
\begin{pmatrix}
I_\text{in} \\
I_\text{out}
\end{pmatrix} = 	\ii \omega C_3 \,	
\underbrace{\begin{pmatrix}
	-1 &  1 \\
	- 1 & 1
	\end{pmatrix}}_{= J_\text{INIC}}
\begin{pmatrix}
V_\text{in} \\
V_\text{out}
\end{pmatrix}.
\end{align}
The in- and output currents are related by $I_\text{in} = - I_\text{out}$, meaning, that both currents flow in opposite directions. For this, the OpAmp must introduce a current source or sink in the system, with currents either flowing inwards to $V_0$ or outwards from $V_0$ in both directions. From the viewpoint of $V_\text{in}$, the INIC acts as a negative capacitance $-C_3$, whereas $V_\text{out}$ experiences the positive capacitance $C_3$. The INIC breaks reciprocity, $J_\text{INIC} \neq J_\text{INIC}^\trans$, in an antisymmetric fashion as the off-diagonal elements differ by a minus sign. Its diagonal contribution is proportional to $\sigma_z$. To avoid it, we add a grounding term of $C_0 \equiv 2 C_3$ to the node at $V_\text{in}$ in order to change the diagonal terms to a unit matrix contribution.

The circuit in Fig.~\ref{fig:circuit_schematic_simplified}a is mathematically described by a Laplacian matrix, which in momentum space under PBCs, takes the form
\begin{align}
J_{\omega}(k)  = \ii \omega \Bigg[ 
&\left(C_1 - \frac{1}{\omega^2 L_1}\right) \begin{pmatrix} 1 & -1 \\ -1 & 1 \end{pmatrix}
+ C_2 \begin{pmatrix} 1 & -\ee^{-\ii k} \\ -\ee^{\ii k} & 1 \end{pmatrix}
+ C_3 \begin{pmatrix} -1 & 1 \\ -1 & 1 \end{pmatrix} \nonumber \\
&- \frac{\ii}{\omega R_\text{x}} \begin{pmatrix} 1 & -1 \\ -1 & 1 \end{pmatrix} -\left( \frac{1}{\omega^2 L_0} + \frac{\ii}{\omega R_0}\right)\begin{pmatrix} 1 & 0 \\ 0 & 1 \end{pmatrix} + C_0 \begin{pmatrix} 1 & 0 \\ 0 & 0 \end{pmatrix}
\Bigg]. \label{eq:SC Laplacian}
\end{align}
Additional to the introduced coupling terms, we include grounding terms $1/(\ii \omega \, L_0)$, $1/R_0$ at all nodes as well as $\ii \omega \, C_0$ at sublattice $A$. By omiiting the resistive intracell connection $R_x$ and representing equation \eqref{eq:SC Laplacian} in the Pauli matrix basis, we arrive at the effective circuit Laplacian stated in equation~(1).

The inductive grounding at each node is used as an $A$-$B$ symmetric grounding, which shifts the imaginary part of the admittance in a frequency-dependent fashion. We damp the circuit globally by grounding all nodes with a resistance $R_0$. The damping is needed to stabilize the circuit and eliminate instabilities (self-amplifying energy gains) in the system, that arise due to the INIC as an active element. The resistance $R_0$ shifts the real part of the admittance eigenvalue spectrum globally by $1/R_0$ but leaves the eigenmodes invariant.

\section{Experimental implementation}

A circuit consisting of ten unit cells was realized on a Printed Circuit Board (PCB). The concise values for the circuit components are detailed in Tab.~\ref{tab:Nominal values}. They are chosen to optimize several aspects of the experimental setup: The Lock-In amplifiers, which are employed for the measurement, operate up to an upper frequency bound of 100\,kHz, the choice of specific parameters facilitates a more stable circuit, in the way that it requires less damping to stabilize it and the non-local response is seen more dominantly if the parameters are adjusted to $r = 2 \gamma$.

In the implementation of the INIC, we employ the unity-gain stable operational amplifier LT1363.
In principle, the auxiliary components in the positive and negative feedback loop of the OpAmp can be chosen arbitrarily, provided they are equal. For a real OpAmp with its finite gain bandwidth product, limited stability conditions, and non-zero output impedance, however, a careful stability analysis~\cite{SuppMat} is required to guarantee the stable operation of LT1363 with our capacitive load $C_3$. This does not only concern the frequency range at which the experiment was conducted, but also frequencies in the MHz range to avoid undesired gain resonances in this regime. Considering those aspects we chose $R_\text{a} = \text{0.2}\,\Omega$ and $C_\text{a} = \text{0.94}\,$ $\mu$F. 

As the INIC is an active, non-reciprocal circuit element it pumps energy into the system. This can lead to eigenmodes of the dynamical matrix, which are described by a complex energy eigenvalue with negative imaginary part, such that their magnitude increases exponentially with time. Hence, energy accumulates in the system until the OpAmp shows non-linear saturation effects and discontinues to function properly. In order to prevent this self-sustained energy gains, we add resistors $R_0$ from each node to ground, which consume the desired amount of power. The suited value for $R_0$, which differs between OBC and PBC, should be small enough to avoid instabilities but large enough to avoid oversized damping, which localizes the voltage response and likewise decreases the measurement accuracy. A numerical computation of the eigenfrequencies using the circuit Hamiltonian formalism~\cite{SuppMat} requires $R_0 \leq \text{45.0} \,\Omega$ for PBC and $R_0 \leq \text{66.8} \,\Omega$ for OBC to remove all global divergences. The discrepancy between those values roots in $\min(\reT(J_\text{OBC})) \geq \min(\reT(J_\text{PBC}))$ for the PBC and OBC spectrum, where a negative real part of the admittance is prone to cause global instabilities. In the actual experimental implementation, parasitic resistances help to stabilize the system and further increase the necessary $R_0$ to stabilize the system.
For the spectral measurements we chose $R_0 = \text{20}\,\Omega$, while for the non-local voltage response measurements we used $R_0 = \text{120}\,\Omega$. 

In order to further eliminate undesired parasitic effects in the experiment and increase its accuracy, we design the circuit boards such that the unintended electromagnetic coupling between different wires and inductors is minimized. We add additional shielding around the inductors, use two circuit boards with five unit cells each, and place them on a metallic mesh. The shielding of the inductors changes their nominal values specified in Tab.+\ref{tab:Nominal values} and furthermore alters their frequency dependence. Moreover, we move the supply lines for the OpAmps out of the PCB board plane. In order to keep parametric disorder to a minimum and thereby preserve an approximate translational invariance, all circuit components were precharacterized by a BK Precision 894 LCR-meter and sorted into groups with only 1\% tolerance.

To perform the spectral measurements, a constant AC current is fed into one node of the board, while lock-in amplifiers with a high dynamical range are used to measure the voltage response of the circuit. The current is realized by a voltage source connected to the PCB through a shunt resistance of $R_\text{S} = \text{12.0}\,\Omega$. The PBC results were acquired by feeding an input current to sublattice $A$ and measuring the voltage at all the other nodes, then repeating the same procedure for sublattice $B$. The spectrum was then calculated through Fourier transformation, which is possible due to circuit periodicity.  For the OBC spectra, the current was fed at each node individually and the voltages of all the other nodes were measured. The results of these measurements are used to populate the Green’s function, as the inverse of the Laplacian,
\begin{align}
\nvec{V} = \begin{pmatrix}
G_{1,m} \, I_m \\
\vdots \\
G_{N,m} \, I_m
\end{pmatrix} = \begin{pmatrix}
G_{1,1} & \cdots & G_{1,N} \\
\vdots & \ddots & \vdots \\
G_{N,1} & \cdots & G_{N,N}
\end{pmatrix}
\cdot \begin{pmatrix}
0 \\ \vdots \\ I_m \\ \vdots \\ 0
\end{pmatrix} \qquad \Leftrightarrow \qquad G_{n,m} = \frac{V_n}{I_m}
\end{align}	
where $I_m$ is the excitation current at node $m$, and $V_n$ is the measured voltage at node $n$. Using this construction, the Laplacian $J$, as well as its eigenvalues and eigenvectors can be computed numerically.

\section{Results and Analysis}

\subsection{Full-scale experimental fit of the theoretical model.}\label{subsec:Full-scale experimental fit of the theoretical model}

\begin{figure}
	\centering
	\includegraphics[width=0.7\linewidth]{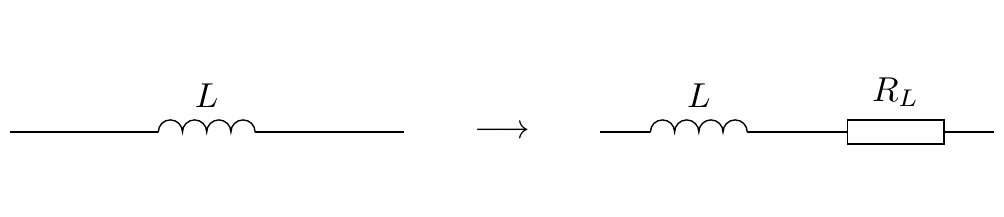}
	\caption{}
	\label{fig:parasitics}
\end{figure}

To fit the experimental results theoretically, we extend our description of the circuit to a full-scale fitted model that takes the most prominent imperfections into account. They occur as parasitic resistances in the inductive couplings of the circuit, and can be modeled as a serial resistor $R_{L_i}$ which modifies the inductances according to
$L_i \quad \rightarrow \quad  L_i + \frac{R_{L_i}}{\ii \omega }$,
as depicted in Fig.~\ref{fig:parasitics}. This results in the full-scale fitted model
\begin{align}\label{eq:full_scale_Laplacian}
J_{\omega,\text{fit}}(k) = \ii \omega \, \Bigg[&\left(C_1 + C_2 + \frac{C_0}{2} - \frac{1}{\omega^2 L_0 - \ii \omega R_{L_0} } - \frac{1}{\omega^2 L_1 - \ii \omega R_{L_1} } -\frac{\ii}{\omega R_0}\right) \, \mathbbm{1} \nonumber \\
-&\left(C_1 -\frac{1}{\omega^2 L_1 - \ii \omega R_{L_1} } +C_2 \cos(k)  \right) \, \sigma_x - \Big(C_2 \sin(k) - \ii\, C_3 \Big) \, \sigma_y \nonumber \\
&+\left(\frac{C_0}{2} -C_3\right)\sigma_z\Bigg],
\end{align}
with capacitances given by $C_1 =$ 300\,nF, $C_2 =$ 94\,nF, $C_3 =$ 47\,nF and $C_0 =$ 94\, nF. The serial resistance to ground, which ensures stability, is chosen to be $R_0 = 20 \, \Omega$ for spectral measurements in accordance with the stability condition evaluated in~\cite{SuppMat}. 

\begin{figure}
	\centering
	\includegraphics[width=\linewidth]{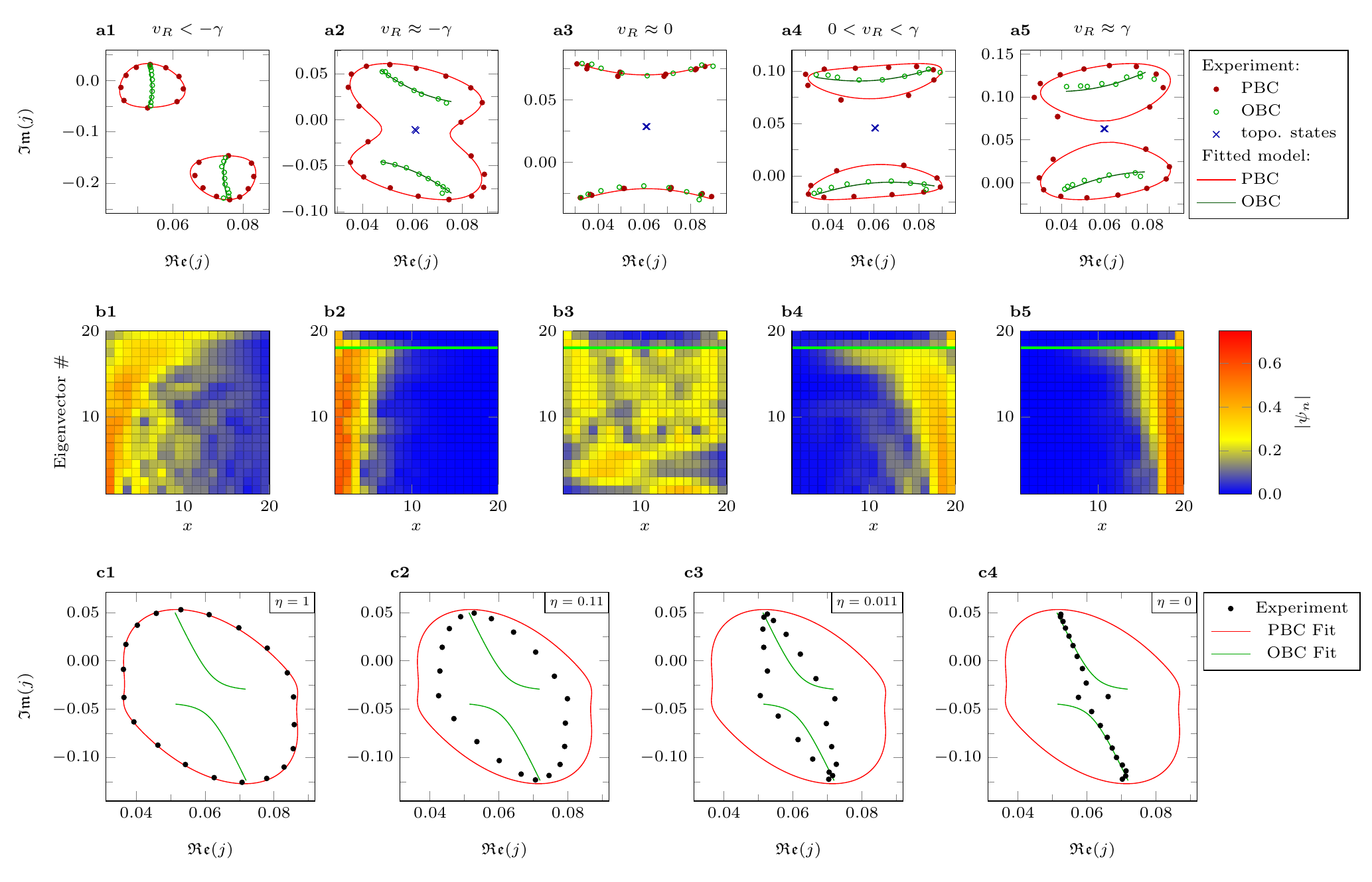}
	\caption{}
	\label{fig:Spectral measurements}
\end{figure}

In comparison to the inductors, the serial resistances of the capacitors are negligible. The filling material of the selected capacitances for the experimental implementation is chosen such that the capacitance is stable across the measured frequency regime. In contrast, the inductors and their corresponding serial resistances $L_1$, $L_0$, $R_{L_1}$, $R_{L_0}$ vary significantly with the AC driving frequency $f=\omega/2\pi$. We fit the PBC and OBC spectra of $J_{\omega,\text{fit}}(k)$ to the measured eigenvalues in Fig.~\ref{fig:Spectral measurements} (a1-a5), and obtain the fit parameters for $L_1$, $L_0$, $R_{L_1}$, and $R_{L_0}$ (see Tab.~\ref{tab:Fit parameters}).

Incorporating the parasitic resistance $R_{L_1}$ in the intracell connection given by $L_1$, the intracell parametric hopping $v(\omega) = C_1 -\frac{1}{\omega^2 \, L_1}$ is adjusted to
\begin{align}
v(\omega) \rightarrow \left(C_1 - \frac{L_1}{\omega^2 \, L_1^2 + R_{L_1}^2}\right) +\  \ii \  \left(-\frac{1}{\omega^2} \cdot\frac{ \, R_{L_1}}{\omega^2 \, L_1^2 +  \, R_{L_1}^2}\right) \equiv v_R+i v_I,
\end{align}
where $v_R$ and $v_I$ label its real and imaginary part. In a theoretical analysis, we are able to connect the limit of periodic and open boundary conditions for $J_{\omega,\text{fit}}(k)$, which differ by a connection with the boundary coupling of $\ii \omega C_2$. To find the full OBC admittance spectrum $j_\text{OBC}$ from the PBC spectrum $j(k)$, we recast the boundary coupling as a pumping of an effective imaginary flux. As detailed in section~\cite{SuppMat}, the OBC spectrum is reached at the smallest, in general finite $\kappa$, where $J(k + i \kappa)$, as an analytic continuation of momentum to the complex plane, recovers the Bloch state limit of plane waves. 

The magnitude of the Bloch factor $\abs{\alpha}$ for the OBC modes relates to the imaginary momentum $\kappa$ by $\abs{\alpha} = \e^{- \kappa}$. For the concrete model, a computation (see~\cite{SuppMat} with $v_R\leftrightarrow v$ and $v_I\leftrightarrow \beta$) yields 
\begin{align}\label{eq:OBC Bloch factor}
\abs{\alpha} = \sqrt{\abs{\frac{v_R+ \ii \, v_I + \gamma}{v_R + \ii \, v_I - \gamma}}} = \sqrt[4]{\frac{(v_R+\gamma)^2+v_I^2}{(v_R-\gamma)^2+v_I^2}}.
\end{align}
The corresponding localization length of the bulk eigenmodes is given by
\begin{align}
\xi = \frac{1}{\kappa} = -\frac{1}{\ln(\abs{\alpha})} = \left[\frac{1}{4} \, \ln(\frac{(v_R-\gamma)^2+v_I^2}{(v_R+\gamma)^2+v_I^2})\right]^{-1},
\end{align}
which translates to the OBC bulk modes through $e^{-x/\xi} = \e^{-\kappa x}$, where $x$ labels the circuit unit cells from left to right. The inverse localization length $\kappa$ is plotted in Fig.~\ref{fig:Phase diagram} as a function of $\omega$. The OBC eigenvalues are correspondingly given by all analytically continued PBC eigenvalues for different momenta $k$,
\begin{align}
j_{\text{OBC}} = \bigcup_k \,  j(k + i \kappa),
\end{align}
where $\kappa = \ln\sqrt[4]{[(v_R-\gamma)^2+v_I^2]/[(v_R+\gamma)^2+v_I^2]}$. For $v_R \neq 0$, the nonzero $\kappa$ yields boundary localized modes for OBC, with the intrinsic localization length of $\xi$.

In Fig.~\ref{fig:Spectral measurements}a, we plot the PBC and OBC admittance spectrum obtained from experimental measurements at differentfrequencies $\omega$. By continuous curves, we show the theoretical band structure obtained from the fitted model including all parasitic effects and circuit fit parameters chosen as detailed in Tab.~\ref{tab:Fit parameters}. For $v_R \neq 0$, we observe an extensive breakdown of bulk-boundary correspondence, as the PBC eigenvalues lying on closed loops coalesce to open arcs for OBC. The corresponding eigenvectors are calculated from the measurement data and displayed in Fig.~\ref{fig:Spectral measurements}b, showcasing the localization of all bulk voltage eigenmodes at either one boundary of the OBC circuit depending on the sign of $v_R(\omega)$. For $v_R <0$ as in Fig.~\ref{fig:Spectral measurements}\,(b1),(b2), all modes are localized at the left edge due to $\kappa > 0 $ ($\xi>0$) as shown in Fig.~\ref{fig:Phase diagram}. The localization length switches its sign as a function of frequency with $v_R >0$ as seen in Fig.~\ref{fig:Phase diagram} and converts the localization of the OBC modes to the right edge in Fig.~\ref{fig:Spectral measurements}\,(b4),(b5).

In our experimental frequency sweep, we cross one transition point, where all modes are delocalized for both PBCs and OBCs as depicted in Fig.~\ref{fig:Spectral measurements}\,(b3). It is found at $v_R= 0$ at 
\begin{align}
\tilde{\omega}_0 = \sqrt{\frac{1}{L_1 \, C_1}- \frac{R_{L_1}}{L_1^2}},
\end{align}
yielding a transition frequency of $\tilde{f}_0 \approx \ $91.3\,kHz. At this point, the localization length $\xi$ for the bulk modes diverges, which is visualized for $\kappa \rightarrow 0$ in Fig.~\ref{fig:Phase diagram}. Bulk-boundary correspondence is restored, leading to a perfect matching of PBC and OBC eigenvalues in Fig.~\ref{fig:Spectral measurements}\,(a3) except for topological SSH eigenvalues. Consequently, the PBC spectrum lies on arcs and follow an extended reciprocity condition, stating that the PBC spectrum is reciprocal around a symmetry point $k_s$ in the Brillouin zone, $j(k_s+k) = j_s(k_s-k)$ (see~\cite{SuppMat}). In the fitted model, the PBC spectrum is symmetric around $k_s = -\arctan(\gamma/v_I) \approx \ $0.4$\pi$ at $\tilde{f}_0$.

For $v_I \neq 0$, due to the incorporation of parasitic resistances, the exceptional points in the OBC spectrum disappear. The strongest localization of the OBC bulk modes is found at the respective global maximum (left edge) and minimum (right edge) of the imaginary momentum $\kappa$, as depicted in Fig.~\ref{fig:Phase diagram}. Due to the small parametric adjustment of $v$ by the incorporation of $R_{L_1}$ as in $v_I(\omega) \ll v_R(\omega)$ and their respective non-trivial frequency dependencies, the frequencies of maximum localization of the OBC modes are marginally adjusted from $\abs{v} = \abs{\gamma}$ to $f_l \approx \ $84.5\,kHz and $f_r \approx \ $100.0\,kHz for the left and right edge localization respectively. Experimental data close to those frequencies is shown in Fig.~\ref{fig:Spectral measurements}\,(a2),(b2) as well as Fig.~\ref{fig:Spectral measurements}\,(a5),(b5), where we observe a strong localization of OBC modes. 

In Fig.~\ref{fig:Spectral measurements}, we further notice the emergence of Su-Schrieffer-Heeger type topological modes, whose eigenvalues reside between the two bulk admittance bands. The topological modes are localized at the boundary of the sample. To find the regime, where topological states exist, we resort to the full-scale fitted model. In analogy to the argument given in~\cite{SuppMat}, a similarity transformation involving the Bloch factor $\alpha$ acting on the OBC Laplacian of $J_\text{fit}$ yields a fully reciprocal model. It presents an intracell hopping of 
\begin{align}
t_0 = \sqrt{(v_R + \ii \,v_I)^2-\gamma^2}
\end{align}
and an intercell hopping of $r$. Two zero-admittance solutions corresponding to boundary-localized topological modes exist, for $\abs{t_0 } < \abs{r}$, which translates to
\begin{align}
\abs{v_R} < \sqrt{\sqrt{r^4-(2 \gamma v_I)^2}-v_I^2+\gamma^2}.
\end{align}
This leads to a frequency regime of 
\begin{align}
f_L < f < f_U,
\end{align}
where we expect zero-admittance solutions in the OBC spectrum and the corresponding topological modes to exist. The zero eigenvalues are shifted according to the frequency-dependent unit matrix contribution in the fitted model. For the given parameters, the topological transition frequencies are given by $f_L \approx \ $78.8\,kHz and $f_U \approx \ $113.0\,kHz. The computed values agree with the measurement, as topological modes occur in Fig.~\ref{fig:Spectral measurements}\,(a2)-(a5), but not in Fig.~\ref{fig:Spectral measurements}\,(a1). Topological modes hence coexist with the localization of all bulk modes for OBC, \ie the non-Hermitian skin effect. Furthermore, the emergence of topological modes is parametrically distorted away from the PBC complex gap-closing point (PBC Exceptional point) due to the non-Hermitian contribution.

We visualize the imaginary flux pumping causing the localization of the OBC modes in an experimental measurement (Fig.~\ref{fig:Spectral measurements}c) supported by a numerical computation (Fig.~\ref{fig:Spectral flow}). The spectral flow of the complex admittance spectrum from PBC to OBC is experimentally achieved through the attenuation of the boundary coupling, that modulates the boundary conditions in Fig.~\ref{fig:Spectral measurements}c.
For PBC the coupling is implemented by the capacitance $C_2$. To realize the imaginary flux pumping, we decrease the capacitance by adding additional capacitors in parallel and, eventually, open the connection completely to achieve OBC. 

In Tab.~\ref{tab:alpha}, we show the theoretically expected magnitude of the OBC Bloch factor $\abs{\alpha(\omega)}$, as computed in equation \eqref{eq:OBC Bloch factor} while using the parameters given in Tab.~\ref{tab:Fit parameters} for the experimentally studied frequencies. The Bloch factor can also be obtained from $\abs{\alpha(\omega)} = \e{-\kappa}$ with $\kappa$ being plotted in Fig.~\ref{fig:Phase diagram}. In comparison, we calculate the mean decay of the OBC bulk eigenmodes, as they are obtained from the measurement data in Fig.~\ref{fig:Spectral measurements}b. The experimental values agree with the theoretical data within the scope of the computed standard deviation. 
This confirms the accuracy of the theoretical fitted model. 
Discrepancies might root in additional parasitic effects and couplings in the realistic setting, which are not included in the fitted model as parasitic resistances of the capacitances, point-like resistive defects due to solder joints or capacitive couplings between conducting wires.

\subsection{Non-local voltage response} 

Consider a linear periodic circuit array with periodic boundary conditions and a single current input at a specific circuit node. Then, the voltage response to the current feed must be localized around the input unit cell, \ie the voltage profile decreases with distance from this unit cell.
In the limit of a dissipationless circuit, the response can also be delocalized but never increase in a different unit cell. 
Otherwise, the system cannot be stable as the voltage accumulates infinitely in a periodic circuit.
Now consider the same circuit with open boundary conditions and a current input in the bulk of the system. If the circuit exhibited bulk-boundary correspondence, the eigenstates of the system's bulk would only change perturbatively and not effect the principal voltage response in comparison to the PBC case. By implication, a non-local voltage response in a periodic circuit is an inevitable signature of BBC breakdown and localized modes of extensive cardinality.

To compute the voltage response to an external current excitation $\nvec{I}$, we resort to the Greens function $G_J = J^{-1}$ of a system with $N$ unit cells. In its spectral representation, it is given by
\begin{align}
G_J = \sum_{n } \frac{1}{j_n} \cdot \frac{\nvec{V}_n \, \nvec{U}_n^\dagger}{\nvec{U}_n^\dagger \, \nvec{V}_n},
\end{align}
where $\nvec{V}_n$ ($\nvec{U}_n$) are the right (left) eigenvectors of the Laplacian matrix to the eigenvalue $j_n$. The voltage response $\nvec{V}$ is then given by $\nvec{V} = G_J\,\nvec{I}$. The eigenstates, which correspond to the smallest eigenvalues are excited the most.
Assume that one eigenvector, say $\nvec{V}_1$, is dominantly excited by the current profile $\nvec{I} = (0,\,\cdots,\,0,\,I_a,\,0,\,\cdots,\,0)^\trans$ with only one nonzero contribution, \ie one input current at node $a$. Then, the voltage response behaves as
\begin{align}\label{eq:non-local voltage response approx}
\nvec{V} \sim \frac{1}{j_1} \cdot \frac{U_{1,a}^*}{(\nvec{U}_1^\dagger \, \nvec{V}_1)} \, \nvec{V}_1 ,
\end{align}
where $U_{1,a} \sim \abs{\alpha}^{-a/2}$ assuming $\abs{\alpha}>1$ (right edge localization w.l.o.g.) as the left and right eigenvectors decay with $\abs{\alpha}^{-1}$ into the bulk. The normalization then follows to be $(\nvec{U}_1^\dagger \, \nvec{V}_1) \sim 2 \cdot N \, \abs{\alpha}^{-N}$ in the circuit system with $N$ unit cells. The voltage profile is approximated by
\begin{align}
\nvec{V} \sim \frac{1}{2 N \, j_1} \cdot \abs{\alpha}^{N- \frac{a}{2}} \cdot \nvec{V}_1.
\end{align}
From this formula, we make two observations. First, the voltage response increases with the nodal distance of the current feed to the localized voltage profile at the right edge. We can therefore associate an amplification factor proportional to $\abs{\alpha}$ to each unit cell in the chain. As we move the excitation further away from the mode localization, the magnitude of the response gets exponentially amplified with $\abs{\alpha}$. Second, the normalization factor $(\nvec{U}_1^\dagger \, \nvec{V}_1)$ decreases as we create a system with larger $N$. A feed on the edge opposite to the localized response is therefore also exponentially increased with the number of unit cells $N$. This makes sense, because the number of intermediate amplification cells is increased in the system. In the limit of $N\rightarrow \infty$, the normalization of left and right eigenmodes vanishes.

Similar to Sec.~\ref{subsec:Full-scale experimental fit of the theoretical model}, we find the mean localization of the OBC bulk eigenmodes, that were obtained from the experimental data to $\abs{\alpha}_\text{nl} = ($2.2 $\pm$ 0.3$)$. It agrees with the theoretically computed value in Tab.~\ref{tab:alpha} within the scope of the standard deviation, but the localization is weakened due to non-linear effects in the operational amplifiers, that become noticeable in the regime of small damping. In Fig.~\ref{fig:non-linear response}a, we show a logarithmic plot of the non-local voltage response with the circuit being excited at node 1 and node 7. It agrees with the expected amplification profile due to the localization of the OBC modes with the Bloch factor of $\abs{\alpha}_\text{nl}$, as shown by a straight line in Fig.~\ref{fig:non-linear response}. Fig.~\ref{fig:non-linear response}b depicts the amplification factor of the circuit with respect to the voltage level at the feed node. The maximum voltage decreases exponentially as a function of the nodal distance to the current feed. The expectation in Fig.~\ref{fig:non-linear response}b using  $\abs{\alpha}_\text{nl}$ as the unit-wise amplification factor agrees with the measurement data. Fig.~\ref{fig:non-linear response} highlights two statements on the non-local response. First, the voltage response retains the decay profile of the localized bulk modes with $\abs{\alpha}_\text{nl}$ and second, $\abs{\alpha}_\text{nl}$ also accurately models the amplification factor as a contribution from each unit cell.

Note, that if we decrease $R_0$ and thereby increase the damping in the system, we need to take the excitation of all modes into account and the approximation formulated in equation \eqref{eq:non-local voltage response approx} breaks down. With larger damping, we introduce interference between all modes in the system leading to a voltage response, that is more and more localized at the point of the current feed.

\begin{figure}
	\centering
	\includegraphics[width=0.5\linewidth]{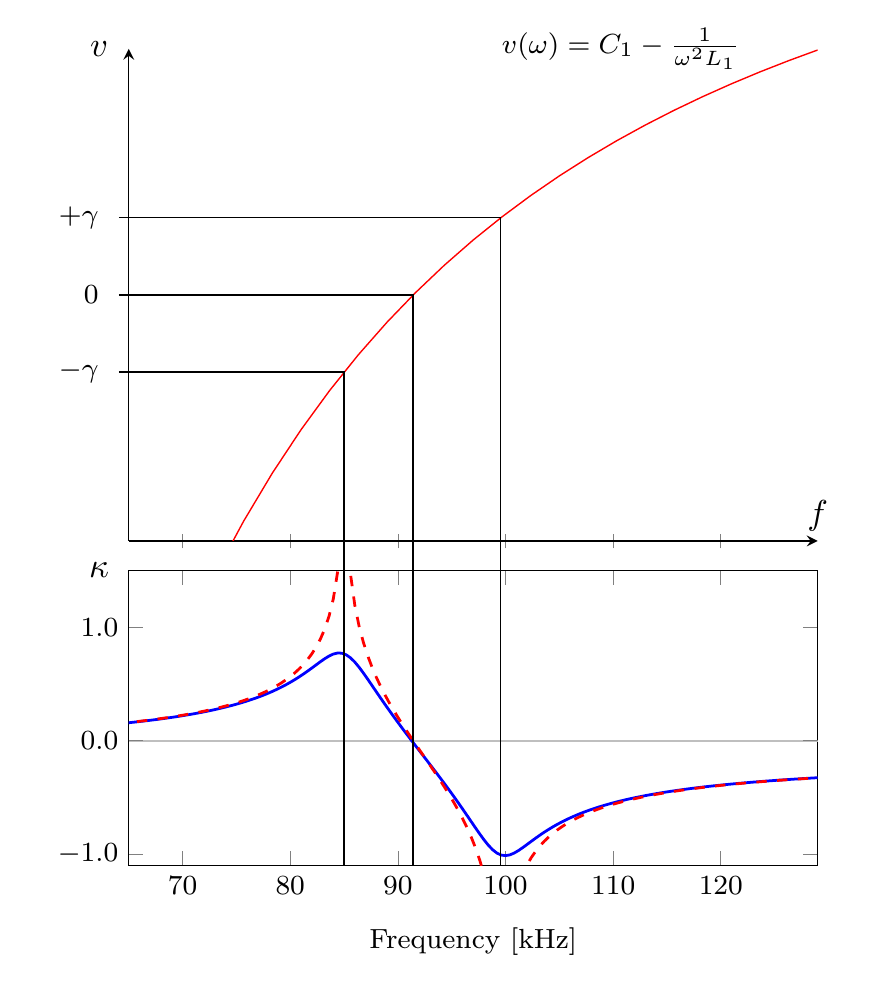}
	\caption{}
	\label{fig:Phase diagram}
\end{figure}

\begin{figure}
	\centering
	\includegraphics[width=\linewidth]{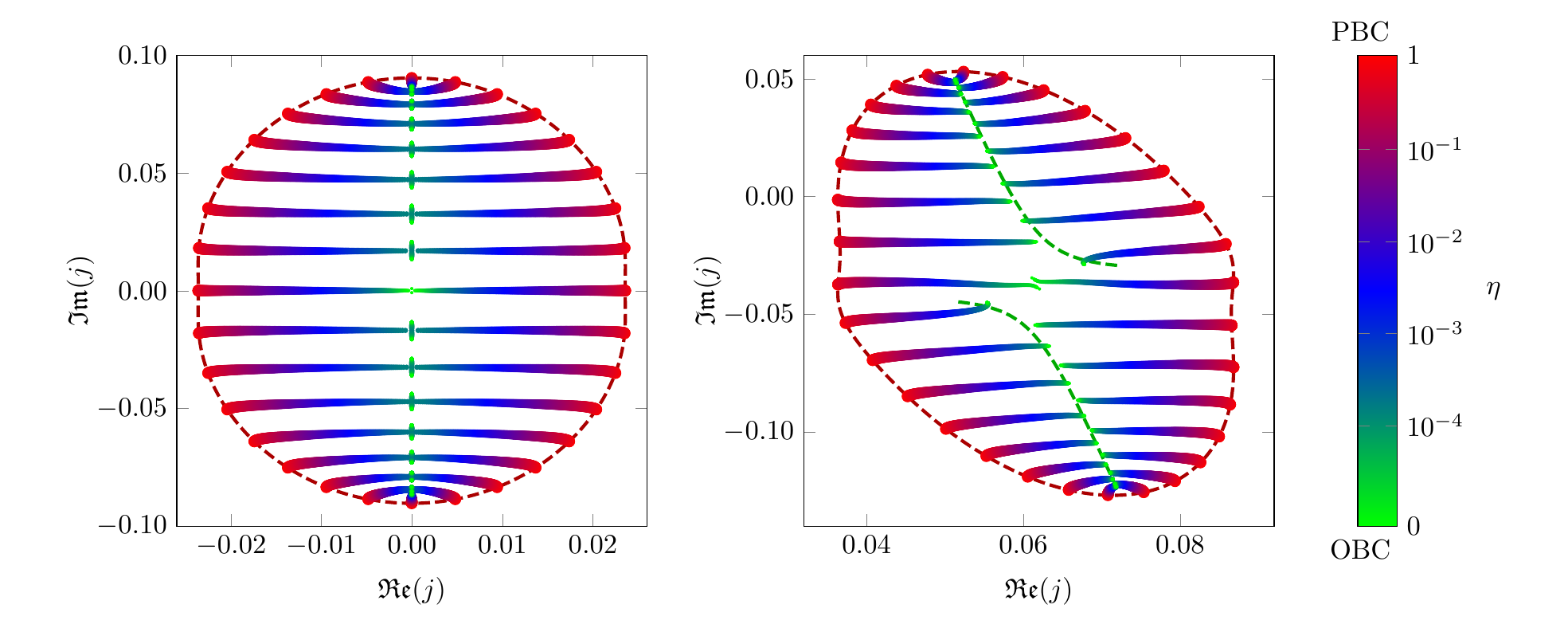}
	\caption{}
	\label{fig:Spectral flow}
\end{figure}

\begin{figure}
	\centering
	\includegraphics[width=\linewidth]{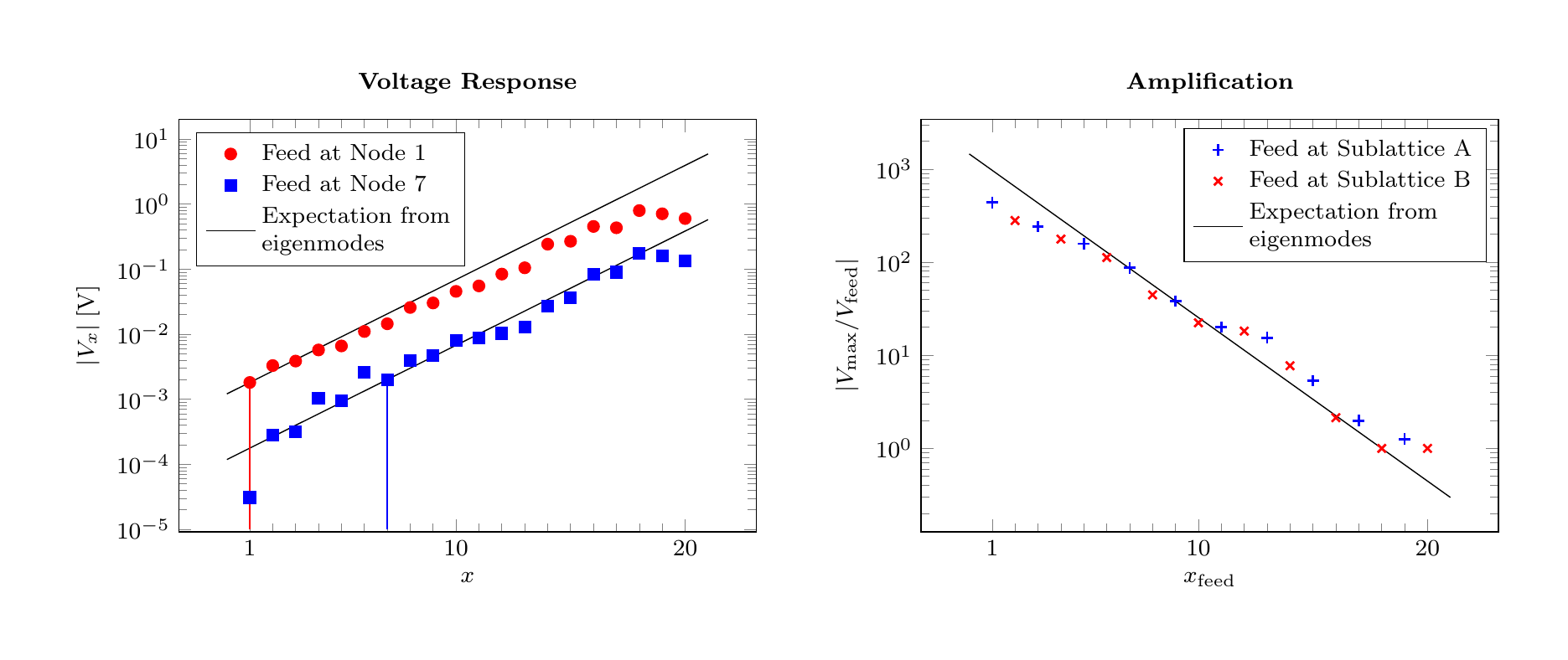}
	\caption{}
	\label{fig:non-linear response}
\end{figure}

Fig.~\ref{fig:Phase diagram} displays the localization of OBC modes, where $\kappa = 0$ means full delocalization. We notice, that the maximum of $\abs{\kappa}$ is larger for right edge localized modes ($\kappa <0$) than for left edge localized ones ($\kappa <0$). This asymmetry is caused by the serial resistances, which exhibit a different frequency dependence than capacitive and inductive components. Their effect is therefore weakened for larger frequencies leading to a stronger localized voltage profile at the right edge than at left edge at the respective frequencies.   

{\bf Caption Table 1:} Nominal values for the components used for the circuit implementation. (${}^*$) The potentiometer implementing $R_0$ was adjusted to $20\,\Omega$ for the measurements of the spectra and to $120\,\Omega$ for the measurement of the non-local voltage response.

{\bf Caption Table 2:} Frequency-dependent fitted inductances and their corresponding serial resistances $L_1, L_0, R_{L_1}, R_{L_0}$ at each experimentally studied frequency. Other experimental parameters are chosen to be $R_0 = 20 \, \Omega$, $C_1 =$ 300\,nF, $C_2 =$ 94\,nF, $C_3 =$ 47\,nF and $C_0 =$ 94\, nF.

{\bf Caption Table 3:} Theoretically calculated and experimentally fitted magnitude of the Bloch factor $\abs{\alpha}$ for the experimentally studied frequencies. Experimental values are obtained as a mean value of the exponential decay profile of all eigenmodes, that have been numerically computed from the measurement data. Their error is given by the standard deviation of the mean value while considering all eigenmodes of the system and therefore quantifies the agreement of the localization of all bulk states.

{\bf Caption Figure 4:} \textbf{(a)} Schematic diagram of one unit cell of the skin circuit including all grounding terms. An $LC$ resonating circuit models the frequency dependent intracell coupling $v(\omega)$. Connectivity between unit cells is modeled by a capacitance $C_2$, which is also depicted in Fig.~1 of the main text. \textbf{(b)} Negative impedance converter with current inversion used in the intracell $A-B$ connection based on an operational amplifier to generate the non-Hermitian, non-reciprocal coupling $\gamma$.

{\bf Caption Figure 5:} Schematic adjustment of the inductive coupling as we incorporate parasitic resistances in the circuit model.

{\bf Caption Figure 6:} \textbf{(a1-5)} Complex admittance spectra for frequencies $f_1=70$\,kHz, $f_2=\ $84.2\,kHz, $f_3= \ $91.5\,kHz, $f_4= \ $95.0\,kHz and $f_5=\ $98.5\,kHz with PBCs in red and OBCs in green. Discrete dots denote measurements, joint lines show the parasitics-corrected theoretical values. \textbf{(b1-5)} Magnitude of voltage modes $\abs{\psi}_n$ for OBCs with $n$ indicating the $n$-th eigenvector of the system, computed from experimental data. The topological modes are separated from the bulk modes by a green line. \textbf{(c)} Spectral flow of the admittance spectrum from PBC to OBC for $f= \ $80\,kHz quantified by $\eta$, which is the boundary bond normalized to its PBC value. Black dots are experimental data points, red and green lines indicate the admittance spectrum of the fitted model for PBC and OBC respectively.

{\bf Caption Figure 7:} Top: Variation of the intracell coupling $v(\omega)$ as a function of frequency. The two OBC exceptional points for the ideal model at $v = \pm \gamma$ and the transition point at $v=0$ with delocalized OBC modes are marked by black lines. Bottom: Localization length $\xi$ of the bulk OBC eigenmodes for the ideal effective model as a dashed red curve and for the full-scale fit model as a blue joint line.

{\bf Caption Figure 8:} Spectral flow of the admittance in the complex plane at the frequency $f = \ $80\,kHz for the effective model (left) and the aberration-corrected fit model (right). The flow is quantified by the parameter $\eta$, which describes the attenuation factor of the 1-20 bond. 

{\bf Caption Figure 9:} \textbf{ (a)} Logarithmic plot of the absolute voltage response in a measurement involving an external current feed with AC driving frequency $f = \ $98.5\,kHz at the left edge of the sample at node 1 in red and at node 4 in blue. The voltage increases exponentially to the right side of the circuit due to a unit-wise amplification. The expected voltage profile normalized to the voltage at the feed node is shown as a straight line, where the slope has been numerically calculated as the average decay profile of the OBC bulk eigenmodes.  \textbf{ (b)} Logarithmic plot of the amplification factor of the circuit defined as the ratio of the maximum voltage $V_\text{max}$ to the voltage  $V_\text{feed}$ at the feed node. The voltage response increases with nodal distance to the right edge, which is supplemented by the expected drop in amplification calculated from the mode localization as in \textbf{ (a)}.

\clearpage

\begin{table}
	\centering
	\begin{tabular}{|c|c|c|c|c|c|c|c|}
		\hline 
		$C_1$ [nF] & $C_2$ [nF] & $C_3$ [nF] & $L_1$ [$\mu$H] & $L_0$ [$\mu$H] & $R_0$ [$\Omega$] & $R_\text{a}$ [$\Omega$] & $C_\text{a}$ [nF] \\ 
		\hline 
		300 & 94 & 47 & 10 & 33 & 20, $120^*$ & 0.2 & 940 \\ 
		\hline
	\end{tabular} 
	\caption{} 
	\label{tab:Nominal values}
\end{table}

\begin{table}
	\centering
	\begin{tabular}{|c|c|c|c|c|c|c|}
		\hline 
		Frequency $f$ [kHz] & 70.0 & 80.0 & 84.2 & 91.5 & 95.0 & 98.5 \\ 
		\hline 
		Inductance $L_1$ [$\mu$H] & 10.1 & 10.1 & 10.1 & 10.1 & 10.1 & 10.1 \\ 
		\hline 
		Inductance $L_0$ [$\mu$H] & 33.0 & 32.0 & 31.7 & 30.3 & 30.1 & 30.0 \\ 
		\hline 
		Serial resistance $R_{L_1}$ [m$\Omega$]& 220 & 230 & 250 & 280 & 300 & 320 \\ 
		\hline 
		Serial resistance $R_{L_0}$ [m$\Omega$]	& 660 & 700 & 750 & 780 & 780 & 780 \\ 
		\hline 
	\end{tabular} 
	\caption{} 
	\label{tab:Fit parameters}
\end{table}

\begin{table}
	\centering
	\begin{tabular}{|c|c|c|c|c|c|}
		\hline 
		frequency $f$ [kHz]	& 70.0 & 84.2 & 91.5 & 95.0 & 98.5 \\ 
		\hline 
		Theory (fit): $\abs{\alpha}_\text{theo}$  & 0.80 &  0.42 & 1.02 & 1.60 & 2.53 \\ 
		\hline 
		Experiment:  $\abs{\alpha}_\text{exp}$ & 0.79 $\pm$ 0.03 & 0.42 $\pm$ 0.03 & 1.0 $\pm$ 0.1 &  1.6 $\pm$ 0.2 & 2.7 $\pm$ 0.3 \\ 
		\hline
	\end{tabular} 
	\caption{} 
	\label{tab:alpha}
\end{table}

\clearpage

\end{document}